\documentclass[12pt]{article}

\usepackage{times}

\usepackage{graphicx}
\usepackage{amsmath,amssymb}

\newcounter{lastnote}
\newenvironment{lastnote}{%
\setcounter{lastnote}{\value{enumiv}}%
\addtocounter{lastnote}{+1}%
\begin{list}%
{[\arabic{lastnote}]}
{\setlength{\leftmargin}{.22in}}
{\setlength{\labelsep}{.5em}}}
{\end{list}}

\title{Aging of dynamically stabilized microtubules}

\author
{Maximilian Ebbinghaus,$^{1,2}$ Ludger Santen$^{1\ast}$\\
\\
\normalsize{$^{1}$Department of Theoretical Physics, Saarland University,}\\
\normalsize{Campus, 66123 Saarbrucken, Germany}\\
\normalsize{$^{2}$Laboratoire de Physique Th\'eorique, Universit\'e Paris-Sud XI,}\\
\normalsize{B\^at. 210, 91405 Orsay Cedex, France}\\
\\
\normalsize{$^\ast$To whom correspondence should be addressed; E-mail:  l.santen@mx.uni-saarland.de.}
}

\date{}

\begin{document} 

\baselineskip24pt

\maketitle

\begin{abstract}
The microtubule network, an important part of the cytoskeleton, is constantly remodeled by alternating phases of growth and shrinkage of individual filaments. Plus-end tracking proteins (+TIPs) interact with the microtubule and in many cases alter its dynamics. While it is established that the prototypal CLIP-170 enhances microtubule stability by increasing rescues, the plus-end tracking mechanism is still under debate. We present a model for microtubule dynamics in which a rescue factor is dynamically added to the filament while growing. As a consequence, the filament shows aging behavior which should be experimentally accessible and thus allow one to exclude some hypothesized models of the inclusion of rescue factors at the microtubule plus end. Additionally, we show the strong influence of the cell geometry on the quantitative results.
\end{abstract}

Microtubules (MTs) are semi-flexible polymers which rapidly switch between a polymerizing and a depolymerizing state, a behavior termed {\it dynamic instability} \cite{Mitchison84}. This feature of the MT dynamics allows the cell to quickly reorganize its MT network in order to adapt to changes in their environment. An important role in MT dynamics is played by the class of microtubule-associated proteins (MAPs) among which the plus-end tracking proteins (+TIPs) have received much attention in past years \cite{Schuyler01,Carvalho03,Galjart03,Akhmanova05,Akhmanova08}. A prototypal protein of these highly conserved +TIPs is CLIP-170, the first protein which was shown to bind to the plus ends of growing MTs \cite{Perez99}.
Typically (e.g. in \cite{Akhmanova08}), two different classes of plus-end tracking mechanisms are presented: First, a mechanism which relies on recognition of a structure at the end of the MT and second, a mechanism in which the +TIP is dynamically included in the filament through addition of new tubulin dimers. As CLIP-170 has been shown to stabilize MTs \cite{Komarova02b}, the distinction between these two scenarios is closely related to the question of whether the MT is dynamically stabilized by addition of CLIP-170 at the growing plus ends or whether the stabilization is due to a structural feature of the filament which does not depend on the MT dynamics. 

In this report, we aim at providing a model which allows one to distinguish between these two stabilization scenarios by their respective aging behavior. For this, we present a stochastic model for MT dynamics whose key ingredients are the consideration of the cell boundary and dynamical stabilization by a rescue factor such as CLIP-170. 
It is known from a previously presented model by Govindan and Spillman \cite{Govindan04} that boundary effects lead to a stationary distribution of MT lengths at a constant and finite value. 
Other scenarios exist which also lead to finite length scales under free conditions \cite{Janulevicius06,Margolin06,Erlenkaemper09}. 
The influence of a stabilizing mechanism was modeled by Antal et al. \cite{Antal07} for an {\it in vitro} scenario in which an unbounded growth of the microtubules is considered. The combination of a confined volume and dynamical stabilization leads to a model which is closer to the {\it in vivo} situation and might give new insights into the stabilization mechanism of MTs by predicting aging of MTs under dynamical stabilization.

The basis of our model relies on the phenomenological description of dynamic instability, 
whose motion in the bulk is usually characterized by four parameters: growth velocity, shortening velocity, rescue (transition from depolymerization to polymerization) frequency and catastrophe (transition from polymerization to depolymerization) frequency \cite{Walker88}. Accordingly, the model consists of a linear filament (MT) of individual subunits (tubulin dimers) which is either in the growing or the shrinking state. Depending on its state, a tubulin dimer of length $\delta$ is added to or removed from the tip of the filament such that the growth resp. shortening rates $\nu_g$ and $\nu_s$ are obtained. Switching of states is stochastic with rescue (shrinking $\to$ growing) frequency $\nu_r$ and catastrophe (growing $\to$ shrinking) frequency $\nu_c$ (Fig. 1).

The model cell has a predefined geometry, an ellipse in our case, within which the origin of nucleating MTs (MT organizing center = MTOC) is chosen. 
An MT grows in a random direction from the MTOC which determines the maximum distance to the boundary $l^*$.
The filament follows the bulk dynamics until it reaches the boundary where switching to the shrinking state is induced with a probability $p_{ind}$. This probability describes experimental observations \cite{Komarova02b} which indicate that an MT filament pauses at the boundary, but that these pauses are too short to be caused by the stochastic switching which is characteristic for the dynamic instability.
An ensemble of such filaments is used to determine the stationary distribution of filament states.

The influence of a rescue factor like CLIP-170 is included in the model in the following way assuming dynamic inclusion of CLIP-170 at the MT plus-end \cite{Carvalho03}. A tubulin dimer that is added to the tip of a growing filament carries a CLIP-170 molecule with probability $p_a$. CLIP-170 molecules on the MT dissociate with rate $\nu_d$ leaving behind a bare tubulin dimer. As for the filament dynamics, the presence of CLIP-170 only affects the rescue frequency \cite{Komarova02b}. Thus, if the MT tip is decorated with a CLIP, a rescue occurs with frequency $\tilde{\nu}_r$ ($>\nu_r$). A formal definition can be found in the supplementary material (SupMat).

The microtubule dynamics is determined by eight parameters. Six of them can be deduced directly from the results in \cite{Komarova02b} (SupMat). The remaining two - rescue frequency in presence of CLIP-170 $\tilde{\nu}_r$ and pre-association probability of a free tubulin dimer with CLIP-170 $p_a$ - determine the experimentally observed rescue frequency in presence of CLIP-170. Knowing the binding stoichiometry of CLIP-170 to newly polymerized MT tips, the rescue frequency $\tilde{\nu}_r$ can also be determined. Throughout this work, $p_a=1$ was chosen and $\tilde{\nu}_r$ was chosen accordingly in order to obtain the correct observed overall rescue frequencies under the stabilizing influence of CLIP-170.

Since, we are particularly interested in the effects CLIP-170 association has on MTs, we also examine the case $p_a=0$, i.e., absence of CLIP-170. In this case, the model can easily be solved exactly in a similar way to \cite{Govindan04} (SupMat). For $p_a>0$, the system exhibits spatial disorder which renders an analytical treatment very challenging. Instead, we performed extensive Monte Carlo simulations to obtain statistics of a large number of microtubules over their entire lifetime.

The choice $p_a=1$ leads to persistent growth up to the cell boundary where frequent switching between growth and shortening is observed (Fig. 2A) before a filament eventually depolymerizes completely. Typical lifetimes are in the order of several minutes as observed in nature. If $p_a=0$, i.e., CLIP-170 does not decorate the filament, MTs are very unstable and depolymerize entirely after a short sojourn time at the boundary (Fig. 2B). This behavior has been anticipated as switches from the growing to the shrinking state and vice versa are rare with respect to the growing and shrinking times resulting from the typical distance between MTOC and cell boundary.

Fig. 3 shows the distribution of active plus ends, i.e., MTs that are not pausing (see also SupMat), with respect to the fraction of the cell radius at which they are located. In contrast to the interpretation in \cite{Komarova02a,Komarova02b} and \cite{Govindan04}, our simulations indicate a very slow increase in the bulk with an accumulation of MT ends at the cell boundary (Fig. S2). In an experimental setup the number of short filaments is systematically underestimated, since the tips are located near the MTOC in a dense environment where they can scarcely be resolved \cite{Komarova02a}. These difficulties obviously do not exist in a theoretical model which explains the higher values of short MT lengths in our model compared to the experimental data (Fig. 3A). The qualitative behavior is not affected by this shortcoming of the experimental method. In the absence of CLIP-170, the distribution of active MT ends is almost flat (Fig. 3B) and thus corresponds well with the experimental data.

The difference in behavior of the model in presence or absence of CLIP-170 is not just quantitative. A dynamical stabilization mechanism via CLIP-170 also qualitatively influences the microtubule dynamics. The association of a rescue factor with the growing plus end through pre-association with free tubulin dimers leads to observable aging at the cell boundary. This is because no further CLIP-170 molecules can be added to the MT tip once it has stopped at the boundary while the older ones continuously dissociate from the filament. This reduces the probability of rescue after a boundary-induced catastrophe. In consequence, the MT ages while being at the boundary, resulting in a shorter remaining lifetime. In order to visualize this, the plot in Fig. 4A shows the probability that an MT filament has not yet completely depolymerized at $t-t_{N}$ seconds after hitting the boundary for the $N$-th time at the time $t_N$, where ``hitting the boundary'' means the contact between filament and cell boundary after an excursion away from the cell boundary which is larger than the experimental threshold. Younger MTs, i.e., those with smaller $N$, have at all times higher survival probabilities. We chose the number of boundary hits instead of the time after the first contact with the boundary as longer excursions away from the boundary and back to it occur. During these excursions, the filament does not age, since new CLIP-170 molecules are added to the tip.
In contrast to these results, the absence of CLIP-170 eliminates the aging behavior which means that all MTs show the very same behavior regardless of the time spent at the boundary. This is portrayed in Fig. 4B which shows identical curves for different values of $N$. 
Depending on the parameters of the CLIP-170 dynamics, the aging effect can be much stronger (SupMat \& Fig. S3).

Cytoskeletal dynamics constitute one of the fundamental processes in the cell which shows generic non-equilibrium behavior. Here, we focus on the dynamics of stabilized MTs, in which case boundary-induced catastrophes are frequently observed. Our results indicate that the actual results are strongly influenced by the cell geometry. In particular, the number of MTs in contact with the cell membrane is drastically increased such that the ability of interaction between the cytoskeleton and the cell membrane is enhanced.

The considered mechanism of dynamic inclusion of CLIP-170 is supported by recent experimental results for animal cells \cite{Bieling08,Dragestein08} whereas in the case of fission yeast, the CLIP-170 homologue Tip1 has been shown to be transported to the MT plus-end by the kinesin homologue Tea2 \cite{Bieling07}. 
Regardless of the exact nature of the MT plus-end tracking of CLIP-170, we have shown that dynamical stabilization of the MT filament will lead to aging phenomena. In contrast to this model of dynamic addition of CLIP-170 molecules to the MT, models in which plus-end tracking is assured by the recognition of a structural feature at the plus-end would not lead to observable aging as recruitment of new CLIP-170 molecules would happen for as long as a plus-end exists and does not depend on addition of tubulin subunits. The existence of aging thus allows one to discriminate between these scenarios \cite{GTPNote}. Our results show that the dynamical interaction leads to destabilization of static filaments. In this way, the cell is provided with the ability to adapt its shape to the environment.

To summarize, the model offers a possibility to experimentally confirm whether the inclusion of a stability factor is mediated by a dynamical interaction with the filament or a structural recognition of the MT plus-end. Namely, the macroscopic effect of microscopic aging should manifest itself in a reduced remaining lifetime of the filament. This quantity is experimentally accessible, sufficient statistics provided.

\bibliography{ebbinghaus}

\bibliographystyle{unsrt}

\begin{lastnote}
\item We thank Y. Komarova for kindly providing the original experimental data, C. Appert-Rolland, M. Evans, R. Harris, and K. Kruse for valuable comments, as well as the DFG Research Training Group GRK 1276 for financial support.
\end{lastnote}

\clearpage

\begin{figure}[tbp]
  \begin{center}
    \includegraphics[scale=0.4, clip]{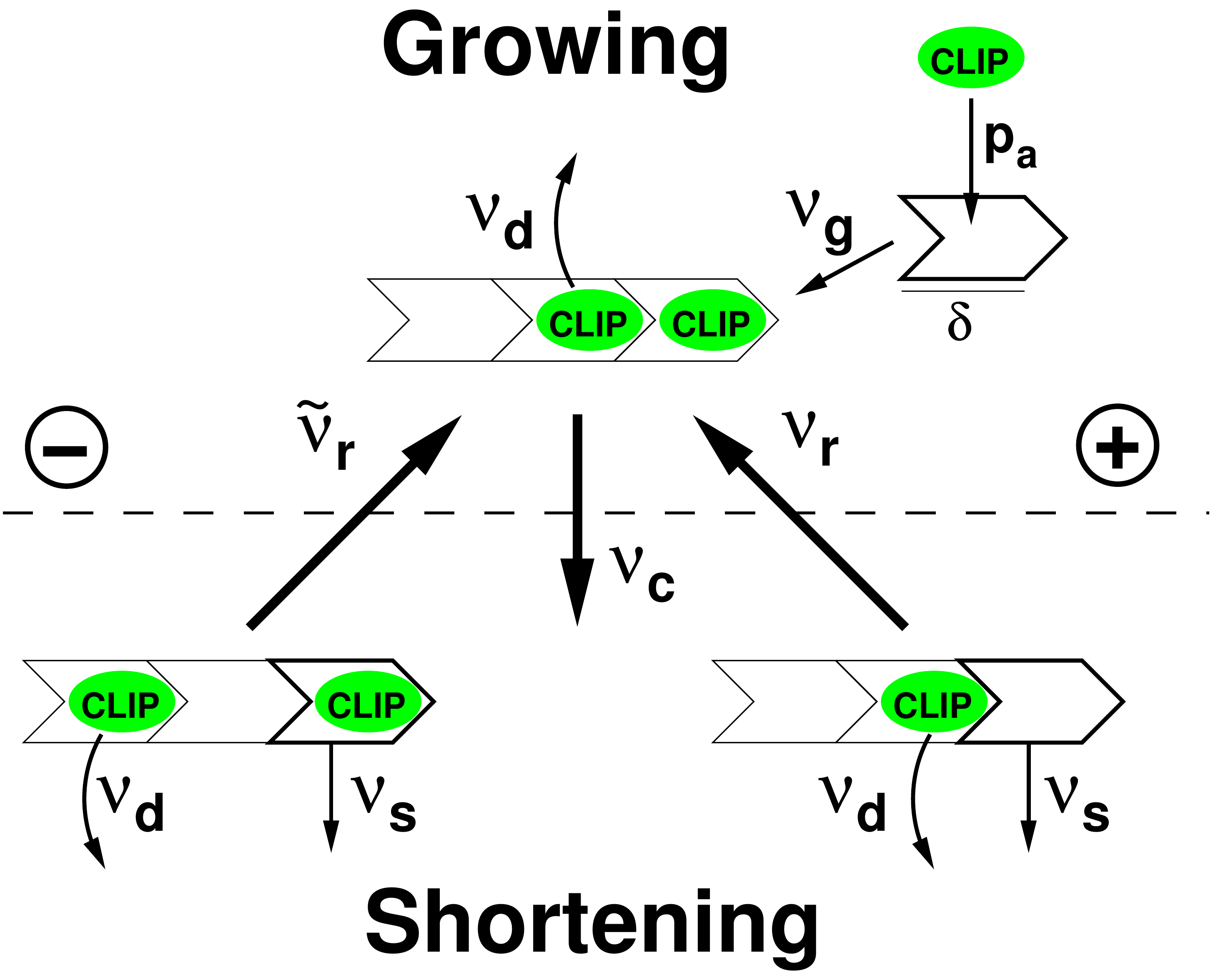}
  \end{center}
  \textbf{Fig. 1. }Schematic representation of the model dynamics far from the boundary. The filament is built from individual subunits of length $\delta$ and can be either in the growing or in the shortening state. In the growing state, a subunit (which is associated to a CLIP-170 molecule with probability $p_a$) is added at a rate $\nu_g$ to the plus end of the filament. Transitions to the shortening state happen with rate $\nu_c$. In this state, the filament loses subunits from its plus end at rate $\nu_s$ regardless of its association to CLIP. However, the switching back to the growing state depends on the presence of CLIP-170 on the last subunit at the plus end and may happen with rate $\tilde\nu_r$ resp. $\nu_r$. At any time, an individual CLIP-170 molecule dissociates at rate $\nu_d$ from the filament.
\end{figure}

\begin{figure}[tbp]
  \begin{center}
    \includegraphics[scale=0.25, clip]{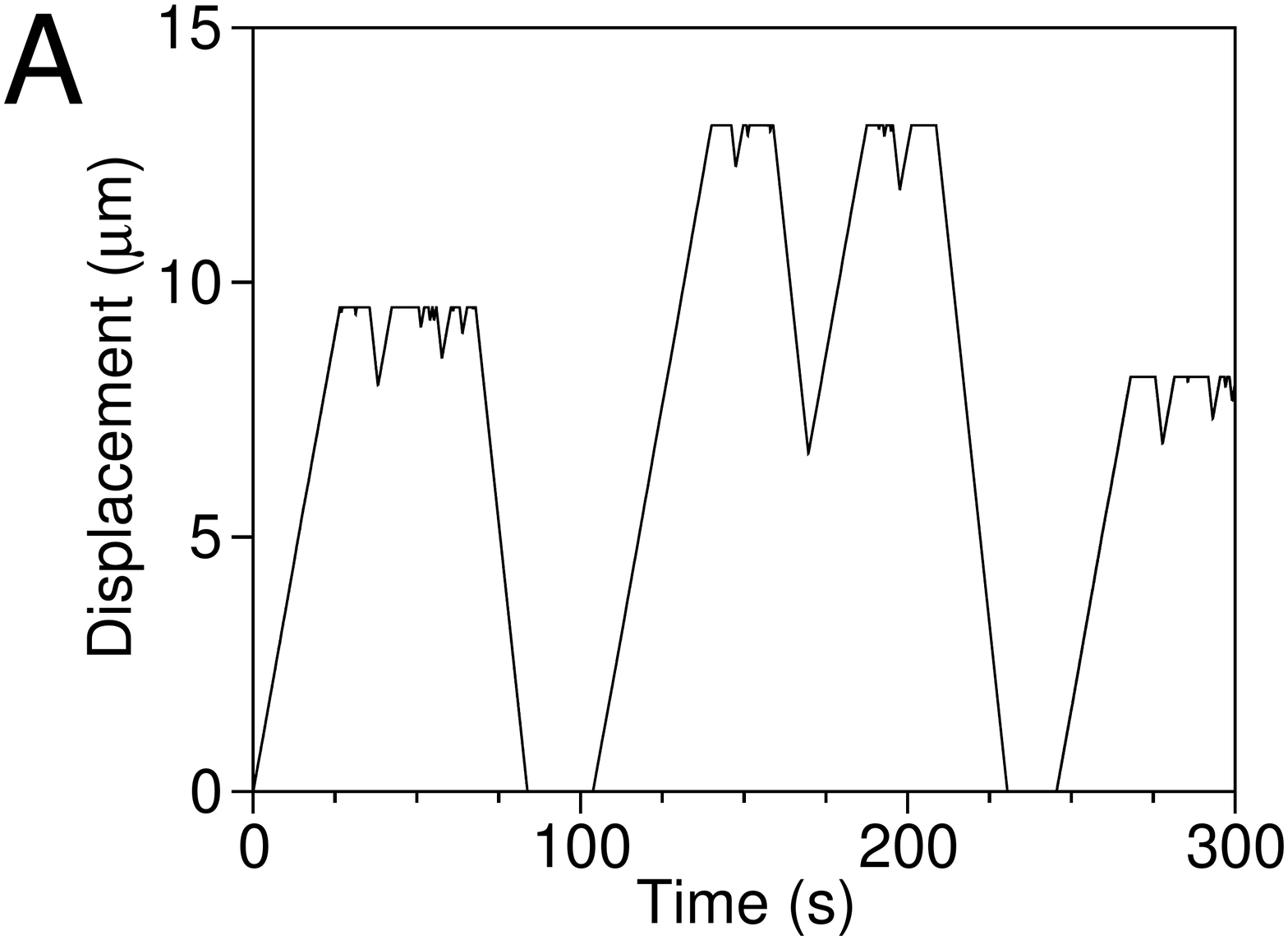}
    \includegraphics[scale=0.25, clip]{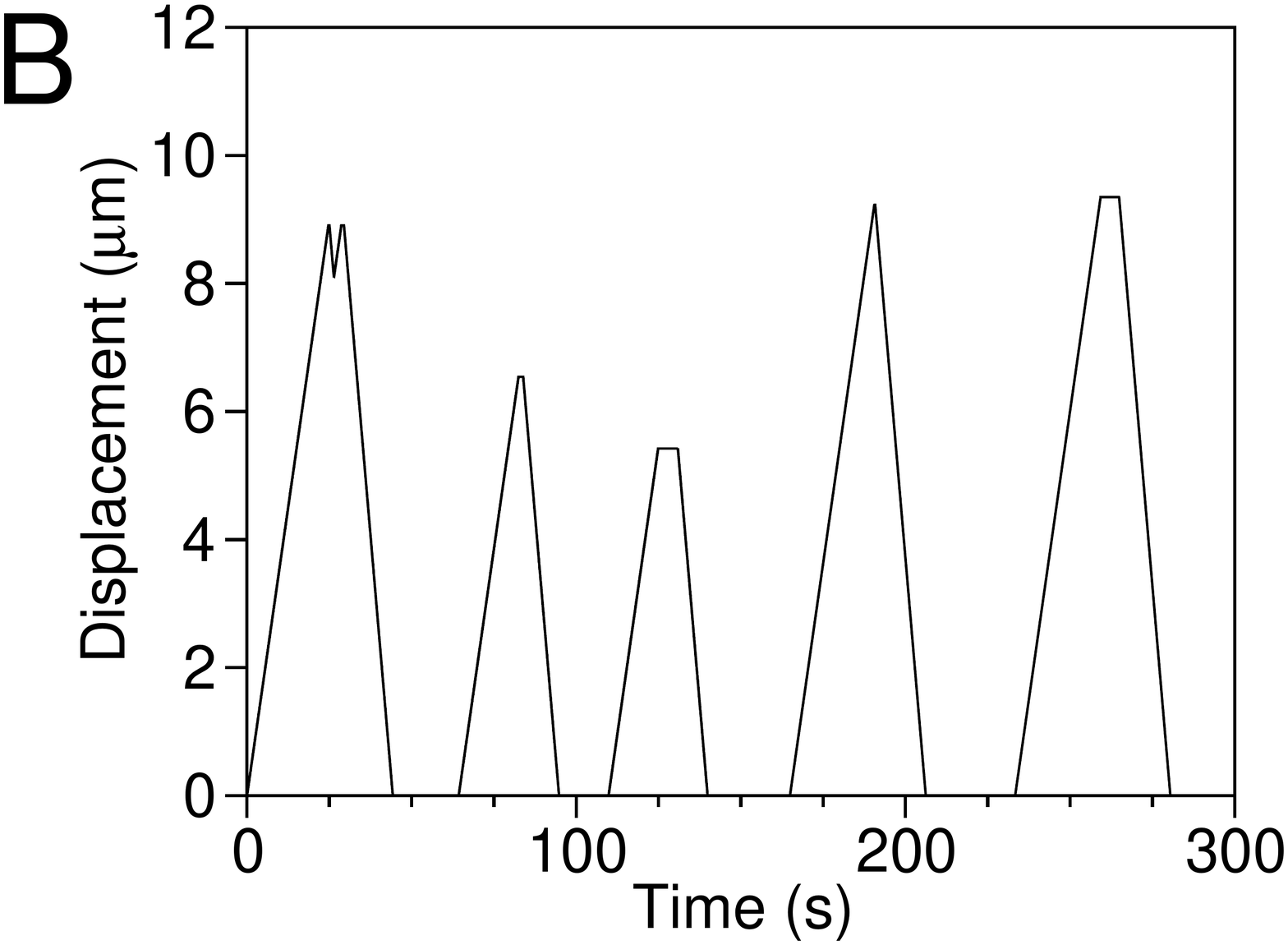}
  \end{center}
  \textbf{Fig. 2. }Life history plots of simulated MTs. (A) In presence of CLIP-170 ($p_a=1$), the MTs show dynamic instability at the cell boundary with frequent rescues. (B) Without the stabilizing effect of CLIP-170, the MTs depolymerize completely after a short contact with the cell boundary. MT life times are mostly determined by the time needed for polymerization to the cell boundary and depolymerization back to the MTOC. The geometry for the simulations was an ellipse of half-axes $a=19.2~\mu m$, $b=6.6~\mu m$ with the MTOC being $1.2~\mu m$ off the center of the ellipse in $x$- and in $y$- direction being representative parameters for the cell geometry discussed in \cite{Komarova02b}.
\end{figure}

\begin{figure}[tbp]
  \begin{center}
    \includegraphics[scale=0.25, clip]{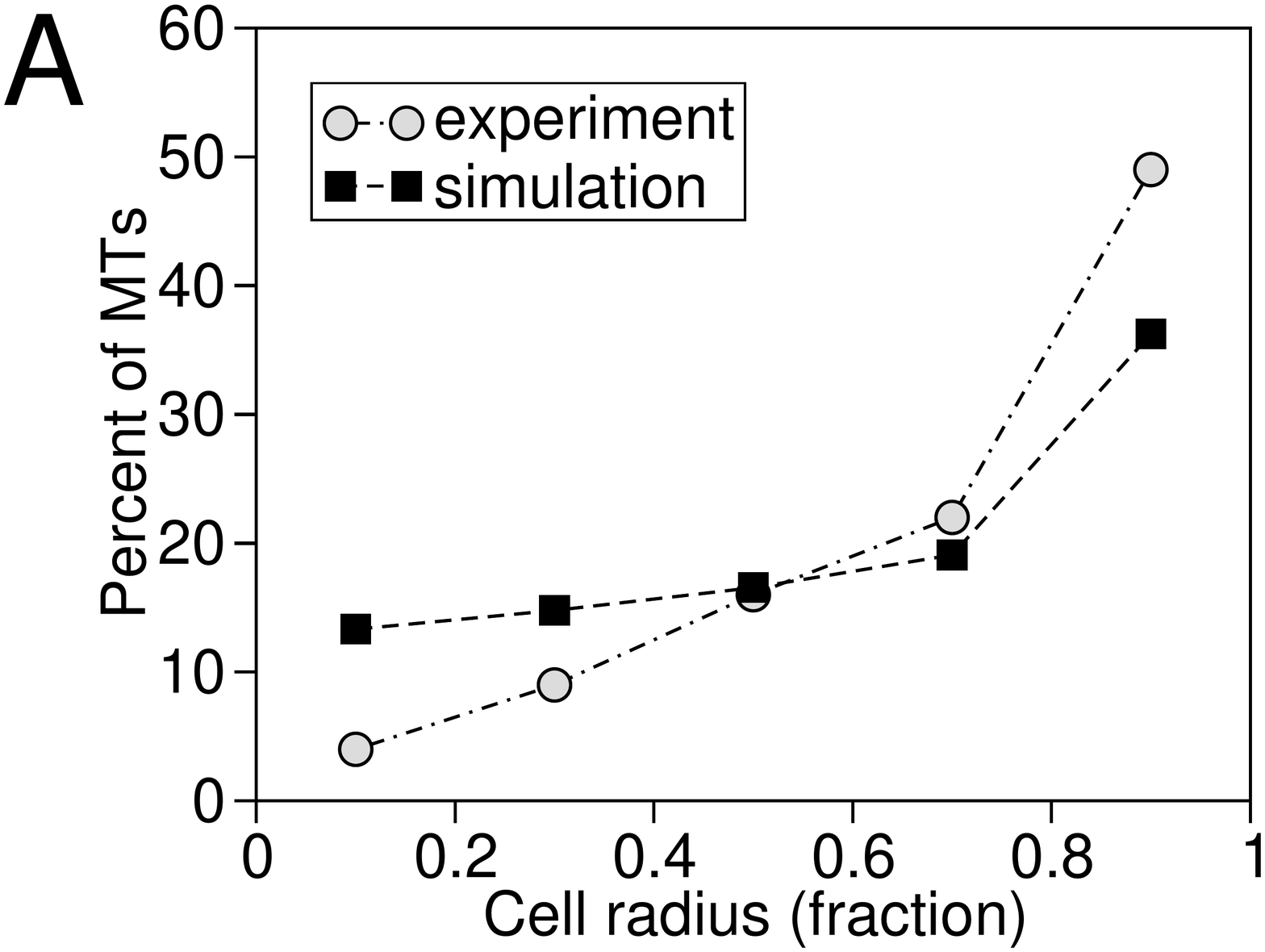}
    \includegraphics[scale=0.25, clip]{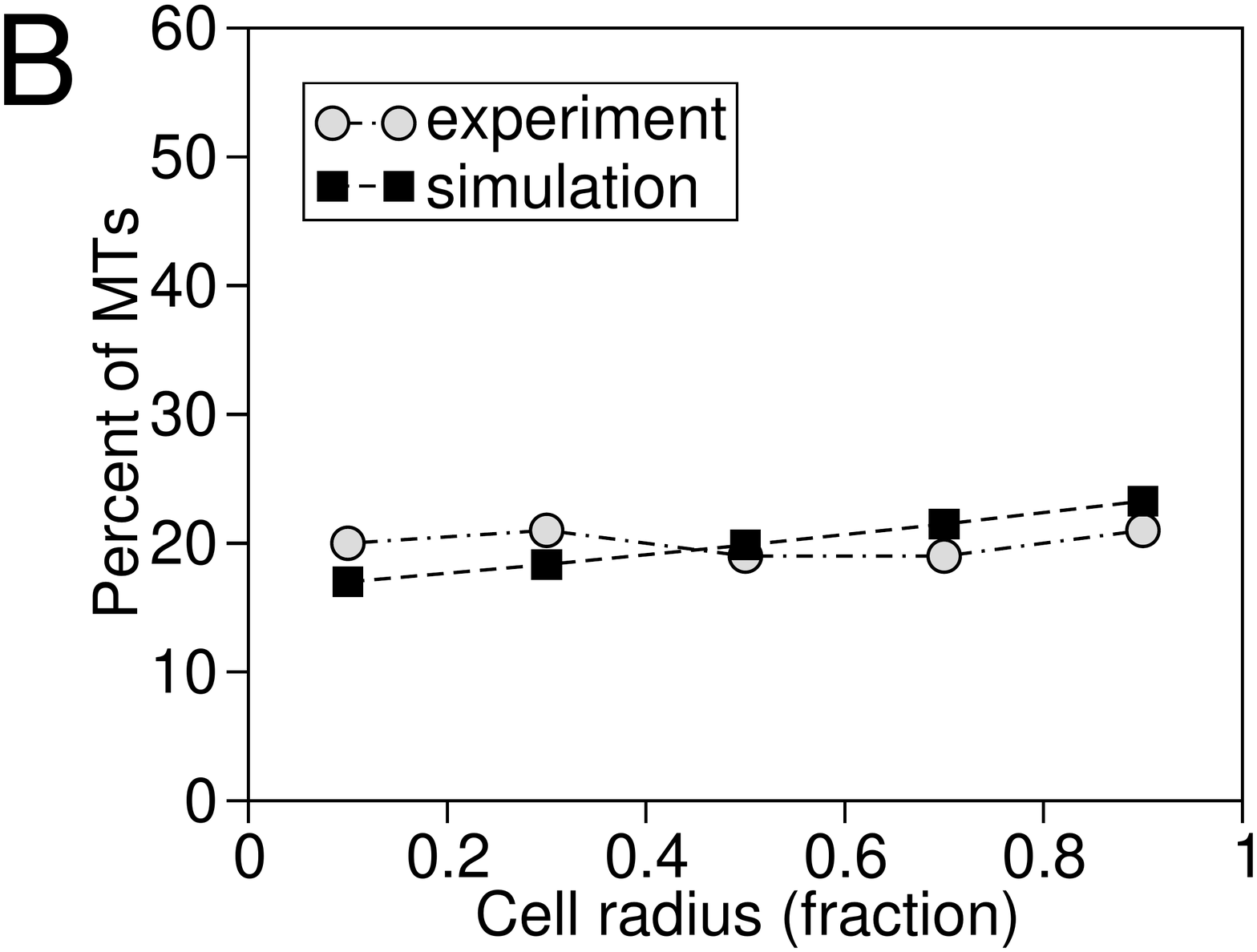}
  \end{center}
  \textbf{Fig. 3. }Histograms of the distribution of active plus ends along the cell radius (comparison with experiment). (A) After correction for paused MTs (definition in SupMat), the simulation results still show an increased number of active MT ends near the boundary in the presence of CLIP-170. (B) In absence of CLIP-170, the distribution of active MT ends increases only very slowly towards the boundary. As in the experiment, MT lengths are rescaled with the cell radius in the corresponding direction of the filament. These dimensionless lengths are robust to variations in cell shape while the absolute ones are not (not shown). Experiments were performed on CHO-K1 and COS-7 cells injected with Cy-3 labeled tubulin \cite{Komarova02b}; the experimental data was provided to us by Y. Komarova.
\end{figure}

\begin{figure}[tbp]
  \begin{center}
    \includegraphics[scale=0.35, clip]{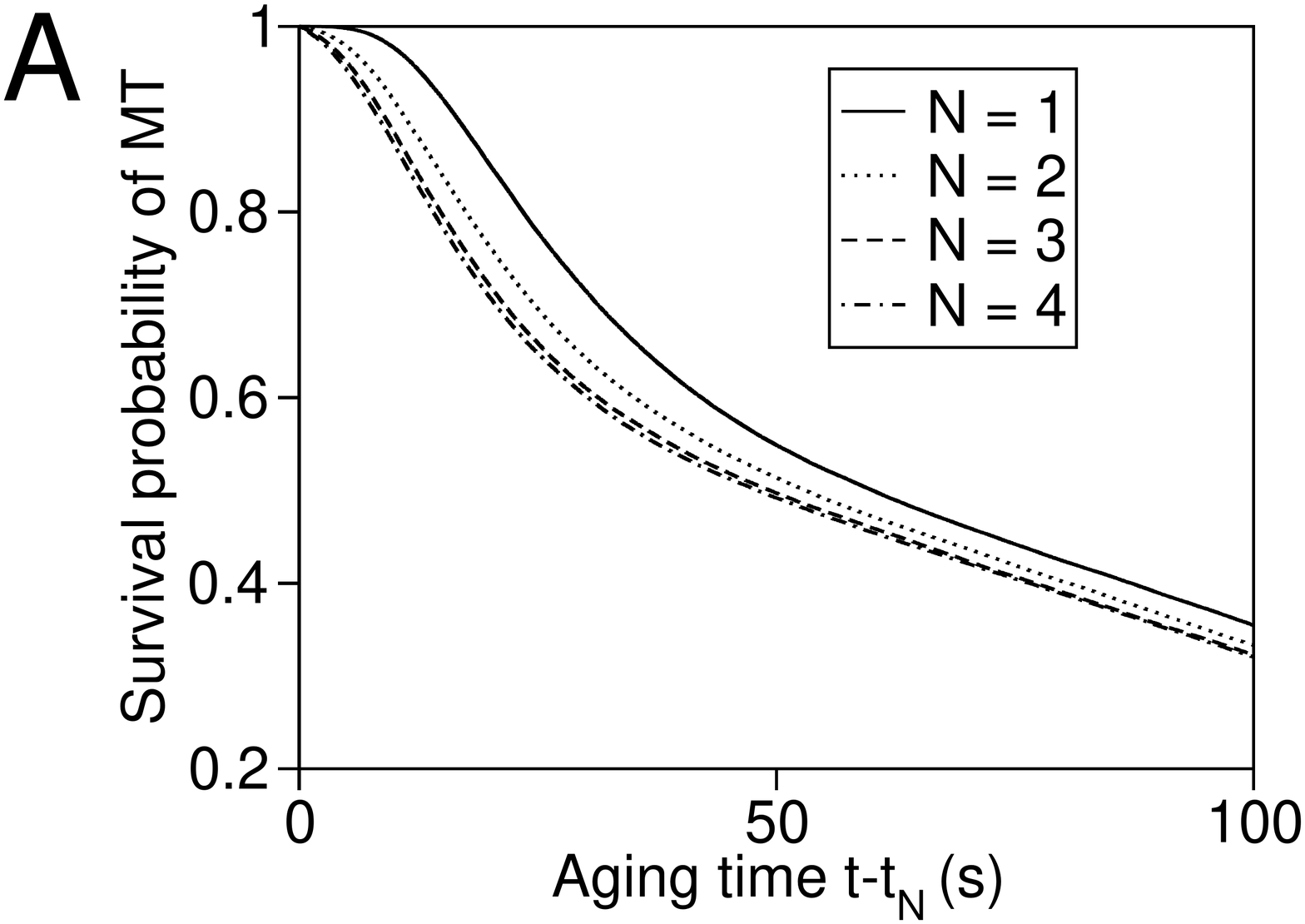}
    \includegraphics[scale=0.35, clip]{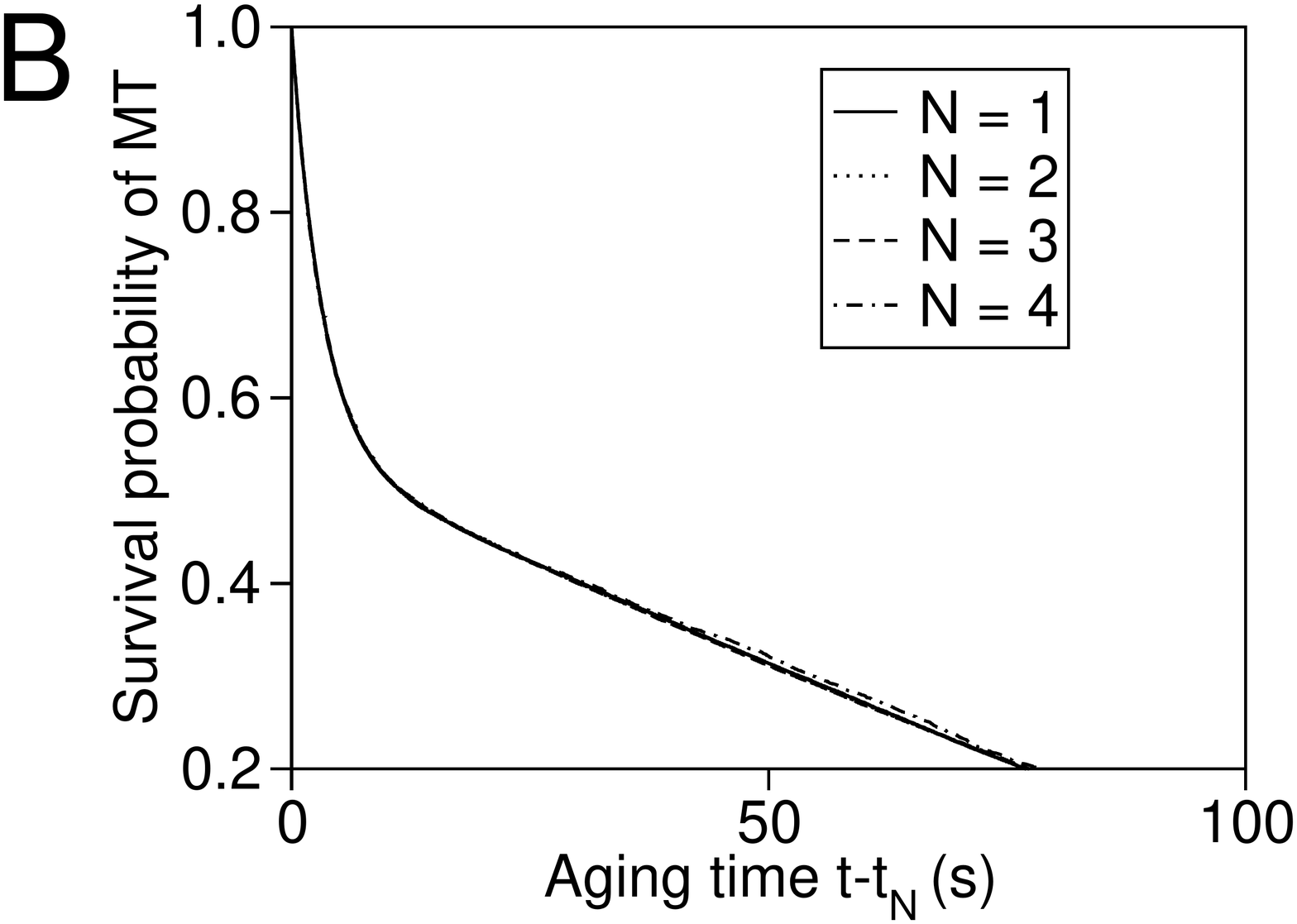}
  \end{center}
  \textbf{Fig. 4. }Survival probability of MT after boundary contact. (A) A dynamically introduced rescue factor like CLIP-170 induces aging of the MT. The survival probability of an MT after it hit the cell boundary for the $N$-th time at the time $t_N$ is lower for higher values of $N$. (B) Without the stabilizing effect of CLIP-170, all MTs show the same behavior regardless of their past, i.e., the number of collisions with the cell boundary. Consequently, all curves collapse.
\end{figure}

\clearpage

\section*{Supplementary Material}

\section{Model}

We consider a filament as a linear arrangement of subunits of length $\delta$ that may or may not carry a CLIP-170 molecule and be either in a growing or a shortening state. Furthermore, the filaments is confined to a limited volume, i.e., it has boundaries to its left and right. By definition, the filament starts growing at the left boundary which is chosen to be at position $l=0$. The distance between left and right boundary determines the maximum length $l^*$.
We represent a configuration of an existing filament in the following way:
\begin{eqnarray}
&&|\ldots --+--+\ldots>\cdots|\\
\text{or}\quad&&|\ldots --+--+\ldots<\cdots|
\end{eqnarray}
The vertical bars represent the system boundaries. The $>$ and $<$ symbols are at the tip of the filament and show the direction of tip motion: $>$ for filament growth and $<$ for filament shortening. The `plus' sign stands for a tubulin subunit which has an associated CLIP-170 molecule, while the `minus' sign illustrates a subunit without CLIP.

The local dynamics are then defined by the following moves.
\begin{enumerate}
\item Dissociation of a CLIP-170 molecule:
\begin{equation}
|\ldots+\ldots\lessgtr\cdots|\qquad\Longrightarrow\qquad|\ldots-\ldots\lessgtr\cdots|\qquad\text{rate }\nu_d
\end{equation}
where $\lessgtr$ is simply a compact way to denote the shortening and the growing state at the same time.
\item Switching from growing to shortening state (catastrophe):
\begin{equation}
|\ldots>\cdots|\qquad\Longrightarrow\qquad|\ldots<\cdots|\qquad\text{rate }\nu_c
\end{equation}
\item Switching from shortening to growing state (rescue) in absence of CLIP-170 at the tip:
\begin{equation}
|\ldots-<\cdots|\qquad\Longrightarrow\qquad|\ldots->\cdots|\qquad\text{rate }\nu_r
\end{equation}
\item Switching from shortening to growing state (rescue) in presence of CLIP-170 at the tip:
\begin{equation}
|\ldots+<\cdots|\qquad\Longrightarrow\qquad|\ldots+>\cdots|\qquad\text{rate }\tilde{\nu}_r
\end{equation}
\item Shortening by one subunit:
\begin{equation}
|\ldots\pm\pm<\cdots|\qquad\Longrightarrow\qquad|\ldots\pm<\quad\cdots|\qquad\text{rate }\nu_s
\end{equation}
with $\pm$ being a subunit either with or without an associated CLIP-170 molecule.
\item Growth by one subunit:
\begin{equation}
|\ldots\pm>\cdots|\qquad\Longrightarrow\qquad\begin{cases}|\ldots\pm+>\cdots|\text{ with prob. }p_a\\
|\ldots\pm->\cdots|\text{ with prob. }1-p_a\end{cases}
\quad\text{rate }\nu_g
\end{equation}
\item At the left boundary, the filament immediately switches to growing state after losing its last subunit:
\begin{equation}
|\pm<\cdots|\qquad\Longrightarrow\qquad|>\cdots|\qquad\text{rate }\nu_s
\end{equation}
\item The right boundary stops filament growth and may induce catastrophe:
\begin{equation}
|\ldots\pm>\quad|\qquad\Longrightarrow\qquad\begin{cases}|\ldots\pm-<|\text{ with prob. }p_{ind}\\
|\ldots\pm>\quad|\text{ with prob. }1-p_{ind}\end{cases}
\quad\text{rate }\nu_g
\end{equation}
\end{enumerate}
The behavior at the boundary, especially the addition of a subunit when undergoing boundary-induced catastrophe, may seem peculiar but was chosen in order to recover exactly the model in [S3] if $p_a=0$ and $p_{ind}=1$. This last site which is not stabilized does not contribute much to the overall behavior in a system of several thousand sites as considered here.

The position of the right boundary is at $l^*$ and depends on the cell geometry that can be chosen arbitrarily. Each filament chooses with equal probability a direction in which it grows from the centrosome. Depending on the cell geometry, the maximum length $l^*$ is then determined.

In accordance with [S3], the fundamental length unit $\delta$ by which a filament grows or shrinks is chosen to be $8~nm/13\approx0.6~nm$, reflecting the length of $8~nm$ of a single tubulin dimer and the fact that an MT typically consists of 13 protofilaments that are arranged in parallel.

\section{Steady state in absence of CLIP-170 ($p_a=0$)}
In the absence of CLIP-170, the steady state probabilities for the filament being in the growing or shortening state with length $l$ can be determined exactly. The master equations that have to be solved are:
\begin{eqnarray*}
\frac{\partial p_+(0,t)}{\partial t}&=&\nu_sp_-(1,t)-\nu_gp_+(0,t)\\
p_-(0,t)&=&0\\
\frac{\partial p_+(l,t)}{\partial t}&=&\nu_gp_+(l-1,t)-\nu_gp_+(l,t)-\nu_cp_+(l,t)+\nu_rp_-(l,t)\quad\text{for }1\leq l < l^*-1\\
\frac{\partial p_-(l,t)}{\partial t}&=&\nu_sp_-(l+1,t)-\nu_sp_-(l,t)+\nu_cp_+(l,t)-\nu_rp_-(l,t)\quad\text{for }1\leq l < l^*\\
\frac{\partial p_+(l^*-1,t)}{\partial t}&=&\nu_gp_+(l^*-2,t)-\nu_gp_{ind}p_+(l^*-1,t)-\nu_cp_+(l^*-1,t)+\nu_rp_-(l^*-1,t)\\
p_+(l^*,t)&=&0\\
\frac{\partial p_-(l^*,t)}{\partial t}&=&\nu_gp_{ind}p_+(l^*-1,t)-\nu_sp_-(l^*,t)
\end{eqnarray*}
where $p_+(l,t)$ resp. $p_-(l,t)$ is the probability for the filament to be in the growing resp. shortening state at time $t$ with tip position $l$.

These equations are only slightly modified compared to those treated in [S3]. Consequently, solving for the steady state solution, i.e., setting the derivatives to zero, can be done straightforwardly in the same way. The solution reads:
\begin{eqnarray*}
p_+(l)&=&\begin{cases}A\cdot a^l\quad&\text{for }0\leq l<l^*-1\\
          A\cdot a^{l^*-1}\left(\frac{1+\frac{\nu_c}{\nu_g}}{p_{ind}+\frac{\nu_c}{\nu_g}}\right)&\text{for }l=l^*-1\\
	  0&\text{for }l=l^*
         \end{cases}\\
p_-(l)&=&\begin{cases}
          0&\text{for }l=0\\
          A\cdot\frac{\nu_g}{\nu_s}a^{l-1}&\text{for }1\leq l<l^*\\
          A\cdot \frac{\nu_g}{\nu_s}a^{l^*-1}p_{ind}\left(\frac{1+\frac{\nu_c}{\nu_g}}{p_{ind}+\frac{\nu_c}{\nu_g}}\right)&\text{for }l=l^*
         \end{cases}
\end{eqnarray*}
with
\begin{equation}
 a=\frac{1+\frac{\nu_r}{\nu_s}}{1+\frac{\nu_c}{\nu_g}}
\end{equation}
and $A$ being a normalization constant such that $\sum_{l=0}^{l^*}[p_+(l)+p_-(l)]=1$. These results have been used in order to check the numerical simulations that serve to produce the results in the spatial disorder case in the presence of CLIP.

\section{Choice of parameters for MC simulations}
Approximate values for the four fundamental parameters of dynamic instability (growth and shortening rate, catastrophe and rescue rate in absence of CLIP-170) can be derived from [S2]. Growth and shortening rates are with respect to the growth resp. shortening of a single subunit of length $\delta$.

The dissociation rate of CLIP-170 can be inferred from the observation that fluorescing CLIP-170 comet tails usually have lengths of 1 to 3~$\mu m$. Assuming a first-order stochastic process, the average time spent on the MT is thus around $6~s$ which is consistent with comet tails that disappear about $5~s$ after MT growth has stopped as observed in [S4].

The probability of boundary-induced catastrophe $p_{ind}$ is difficult to determine as there is little experimental data. Nevertheless, experimental trajectories of MTs without CLIP-170 in figure 2F of [S2] show that MT tips pause at the boundary before undergoing catastrophe. The time spent seems to be in the range of a few seconds, although more experimental data is needed in order to reach a higher precision on this parameter. In combination with the growth rate one obtains the value of $p_{ind}$ shown in table S1.

As stated in the main text, the combination of CLIP-170 association probability $p_a$ and rescue frequency in presence of CLIP-170 $\tilde{\nu}_r$ is determined by choosing $p_a=1$ and adjusting $\tilde{\nu}_r$ in order to obtain the observed overall rescue frequency of $0.17~s^{-1}$ [S2]. This leads to values of $\tilde{\nu}_r$ which are about 100 times bigger than $\nu_r$, but which are still a lower bound for the real value, since it can be assumed that very rapid sequences of catastrophe and rescue were not observable in the experiments even though they are counted in our simulations.

\begin{table}[h]
 \begin{center}
  \begin{tabular}{|l|r|}
\hline
\textbf{Parameter} & \textbf{Simulation value}\\
\hline
Growth rate $\nu_g$ & $600~s^{-1}$\\
Shortening rate $\nu_s$ & $1000~s^{-1}$\\
Catastrophe rate $\nu_c$ & $0.003~s^{-1}$\\
Rescue rate without CLIP-170 $\nu_r$ & $0.024~s^{-1}$\\
CLIP-170 dissociation rate $\nu_d$ & $0.15~s^{-1}$\\
Probability of boundary-induced catastrophe $p_{ind}$ & $5\cdot10^{-4}$\\
\hline
  \end{tabular}
 \end{center}
\textbf{Table S1. } Standard set of parameters derived from experimental data. The remaining parameters are discussed in the main text.
\end{table}

\section{Parameter dependence of aging behavior}

The aging effects observed in Fig. 4A depend mainly on the dissociation rate of CLIP-170 $\nu_d$ and the rescue rate if the tip subunit is occupied by a CLIP-170 molecule $\tilde\nu_r$. In Fig. S3, the survival probability of an MT is plotted for different combinations of these two parameters in order to elucidate their influence on the aging behavior.

At constant CLIP-170 rescue rate $\tilde\nu_r$, the impact of aging becomes stronger if the dissociation rate $\nu_d$ is decreased (Fig. S3 B-D). Moreover, the dissociation rate $\nu_d$ controls the time scale which is relevant for the survival probability.

A variation of the CLIP-170 rescue rate $\tilde\nu_r$ is also able to alter considerably the aging behavior (Fig. S3 A, C, E). For very high values of $\tilde\nu_r$, the number of boundary contacts seems to be irrelevant for the survival probability after the second boundary contact (Fig. S3 E, F).

\section{Influence of geometry}
The cell geometry has a big impact on the certain quantities in the absence of CLIP-170. Hence, the length distribution of shortening episodes (definition below) is mainly an indicator for the shape of the cell rather then for the MT dynamics. Indeed, simulations with different geometries reveal very different distributions of shortening lengths (Fig. S1). Nevertheless, the qualitative behaviour is conserved over a wide range of geometric parameters and rescaled quantities as those presented in Fig. 3 correlate well with experimental results.

\section{Numerical simulations and data treatment}

\subsection{Simulation method}
For the simulation of this model, a standard Monte Carlo algorithm was implemented: The sum of the rates of all possible moves in the current state was calculated before choosing one of these moves via tower sampling [S1]. When the filament depolymerized completely to length zero, the new growth direction was drawn from a uniform distribution, and the resulting maximum filament length was determined and kept constant until the new filament depolymerized again. The cell geometry including the position of the MTOC was kept constant throughout the whole simulation.

\subsection{Distribution of active plus ends}
Active plus ends were defined as MT ends that were either growing or shrinking, similar to [S2]. In our model, pausing filaments can only exist at the cell boundary where MTs have their maximum length if $p_{ind}$ was chosen. In this case, the filament paused before depolymerizing and, consequently, the boundary site at $l^*$ was excluded from the measurements of MT length distributions as filaments at maximum length would not be identified as being active. All other MT lengths were classed into a histogram with five bins that represent 20\% of the maximum length $l^*$ each.

\subsection{Shortening lengths and aging behavior}
Shortening lengths were determined as the distance between the site on which catastrophe occurred and the site on which the filament was rescued. If the filament depolymerized completely, the rescue site was taken to be at the origin. The threshold of experimental measurements was $0.18~\mu m$ [S2] which corresponds to approximately $300 \delta$. Since measurements below this value were not possible, we excluded shortening events of less than 300 subunits. The same applies to the excursions used to determine the aging behavior: If the excursions away from the cell boundary were below the experimentally detectable threshold, the counter for boundary hits was not increased.

\subsection{Catastrophe and rescue frequency}
Whenever an MT switched from growing to shrinking (shrinking to growing), a catastrophe (rescue) event was scored if the event did not take place at one of the boundary sites. The sum was then divided by the total simulated time.

\clearpage

\section{Supplementary figures}

\begin{figure}[h]
  \begin{center}
    \includegraphics[scale=0.25, clip]{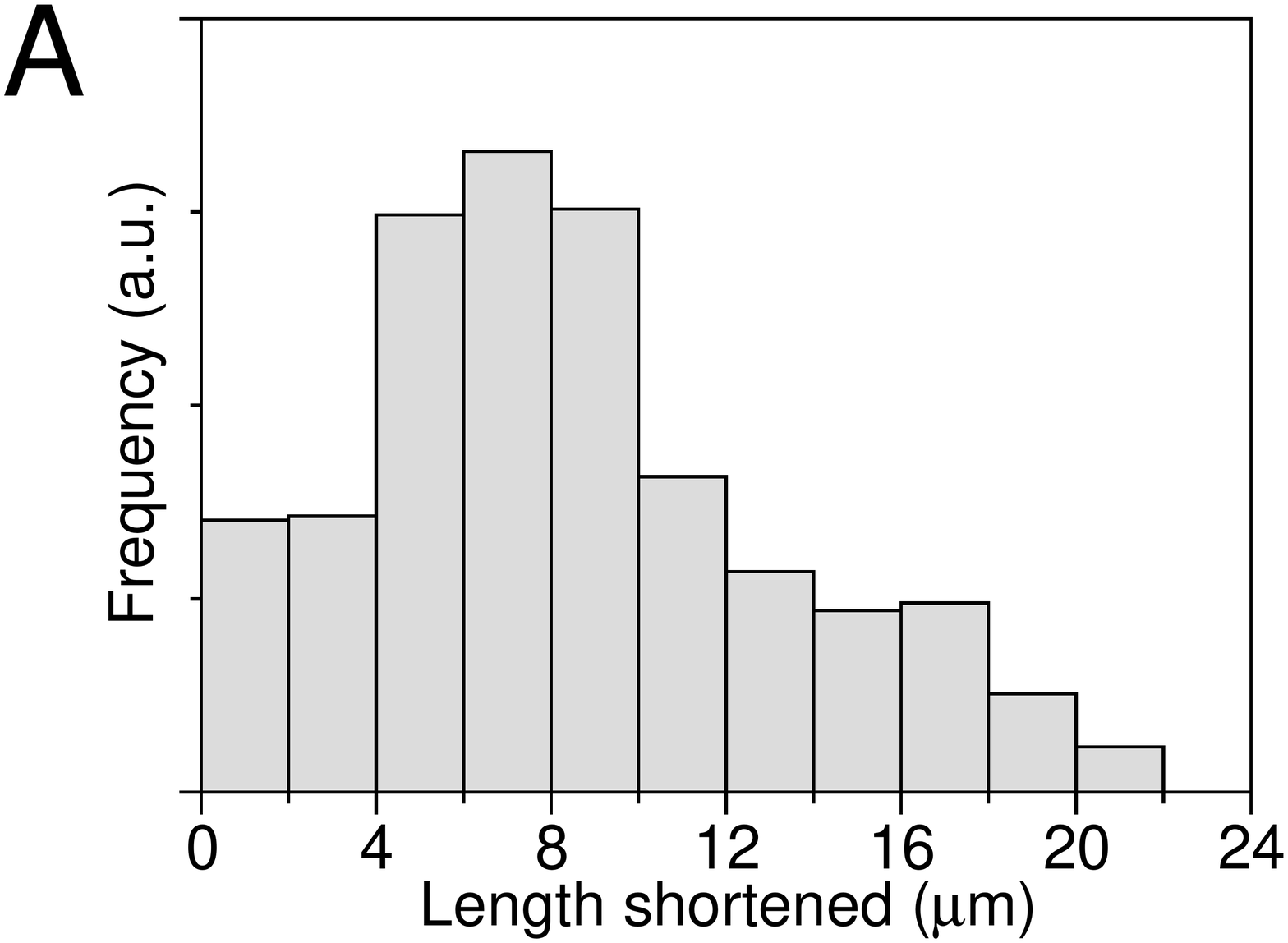}
    \includegraphics[scale=0.25, clip]{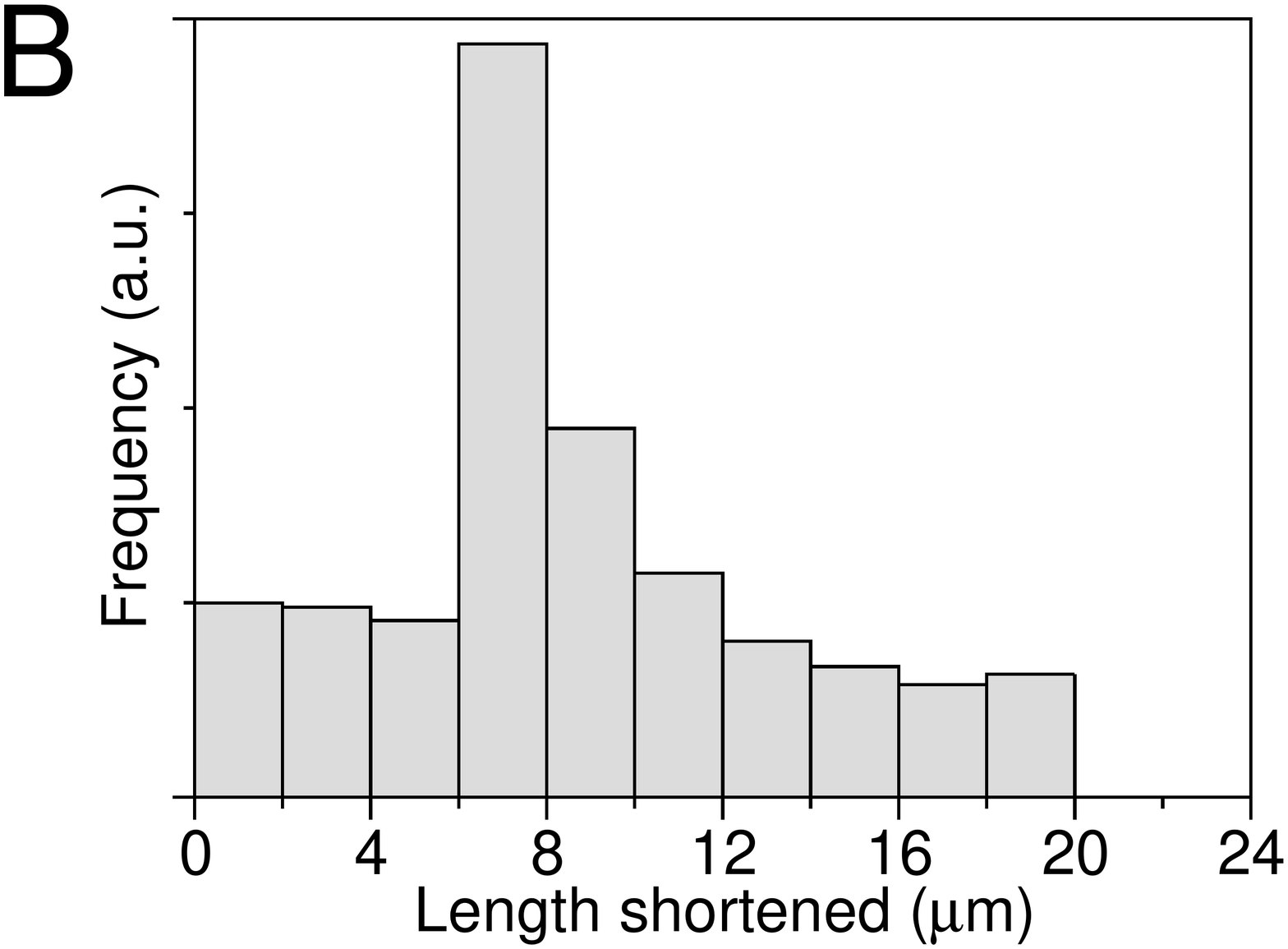}
    \includegraphics[scale=0.25, clip]{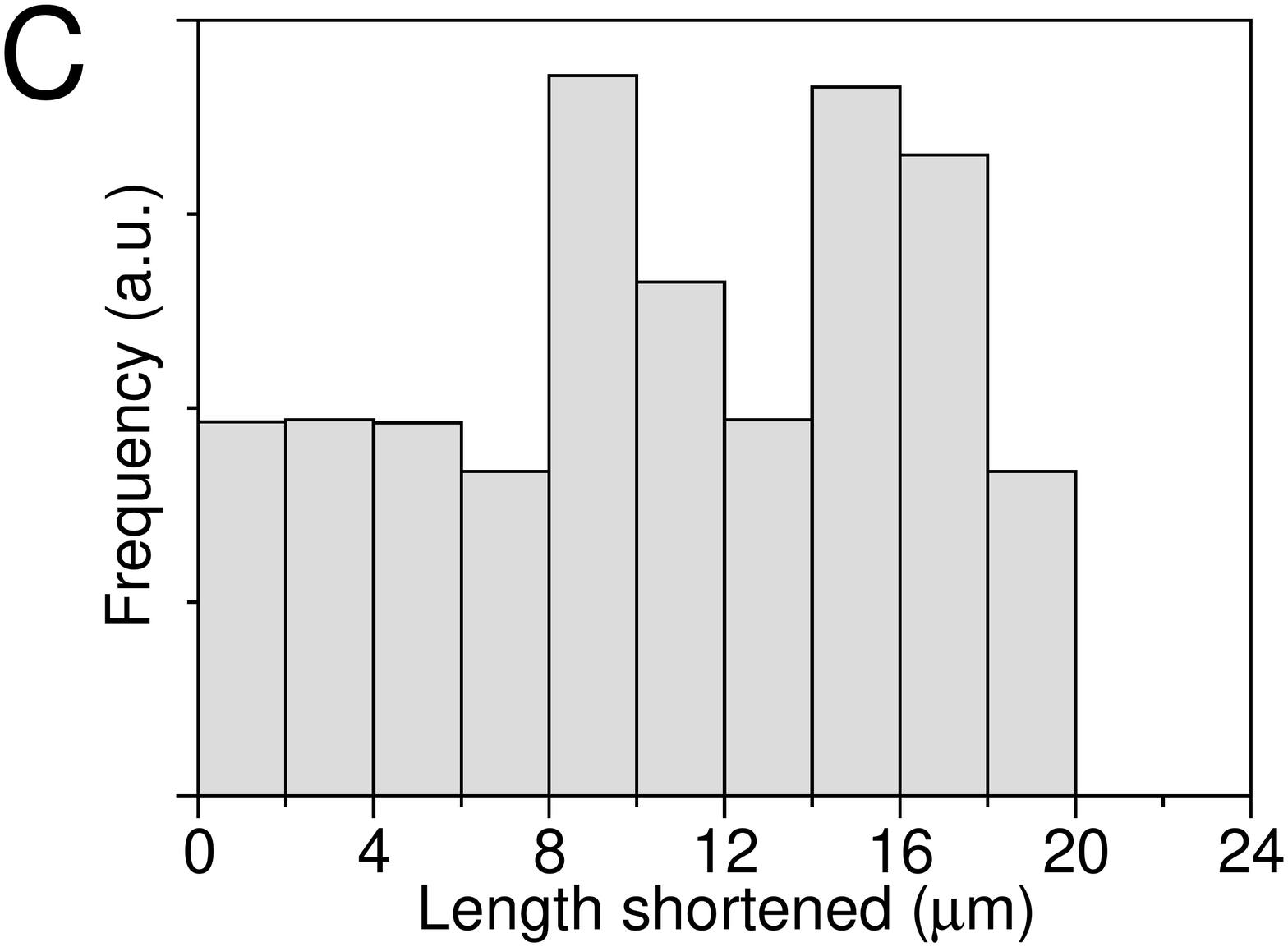}
    \includegraphics[scale=0.25, clip]{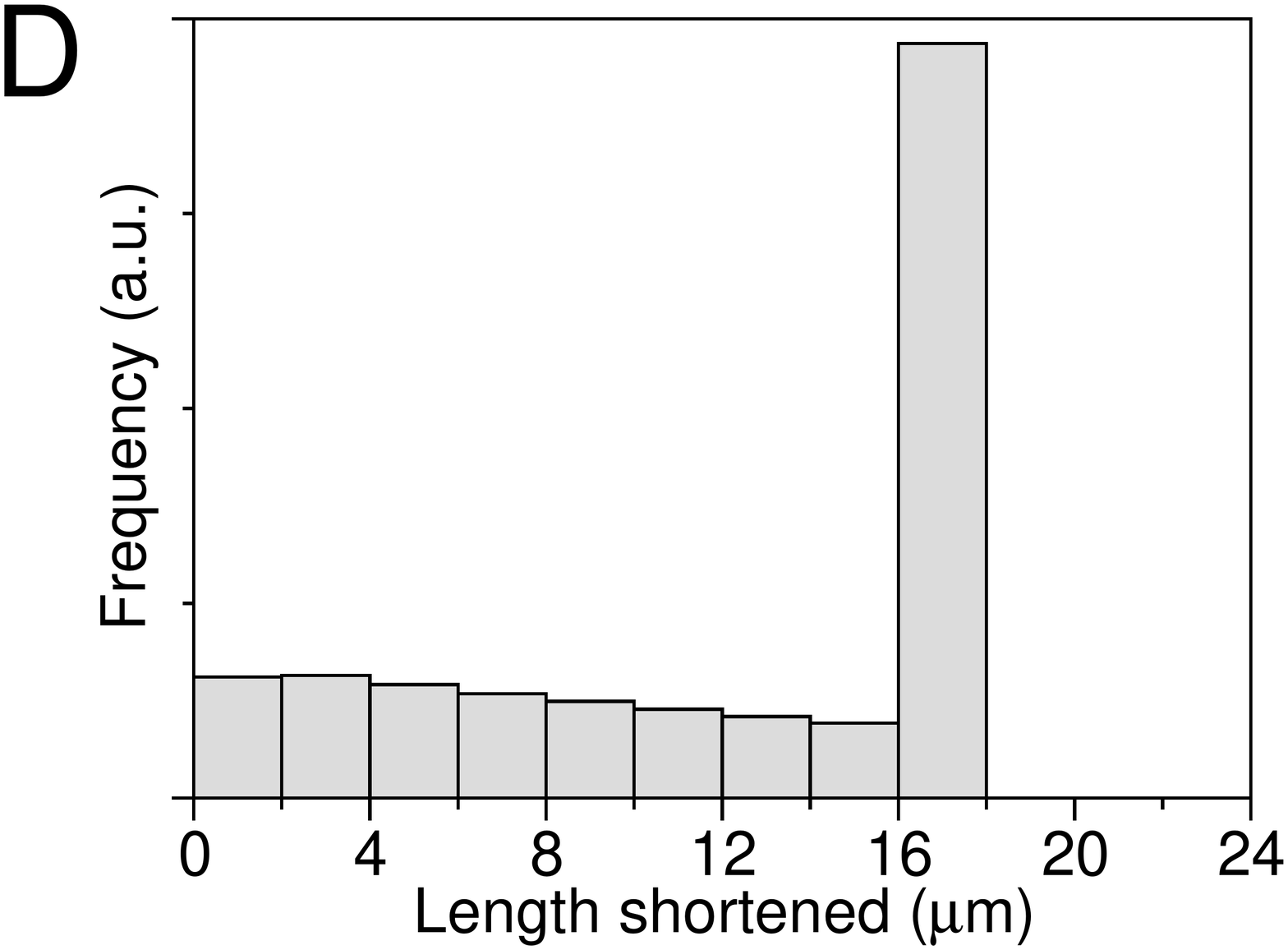}
  \end{center}
\textbf{Fig. S1. }Examples for distributions of shortening lengths in absence of CLIP-170 for different geometries. With the same set of parameters, very different distributions are obtained depending on the geometry. Reproducing the distribution experimentally determined in [S2] is thus a question of finding the correct geometry. Chosen geometries were an ellipse with half-axes $a=19.2~\mu m$, $b=6.6~\mu m$ (MTOC at $(1.2~\mu m, 1.2~\mu m)$ off the center (A) resp. MTOC at the center (B)), an ellipse with half-axes $a=18~\mu m$, $b=12~\mu m$ (MTOC at $(0~\mu m, 3~\mu m)$) (C), and a circle of radius $r=18\mu m$ (MTOC at the center) (D).
\end{figure}

\begin{figure}[h]
  \begin{center}
    \includegraphics[scale=0.4, clip]{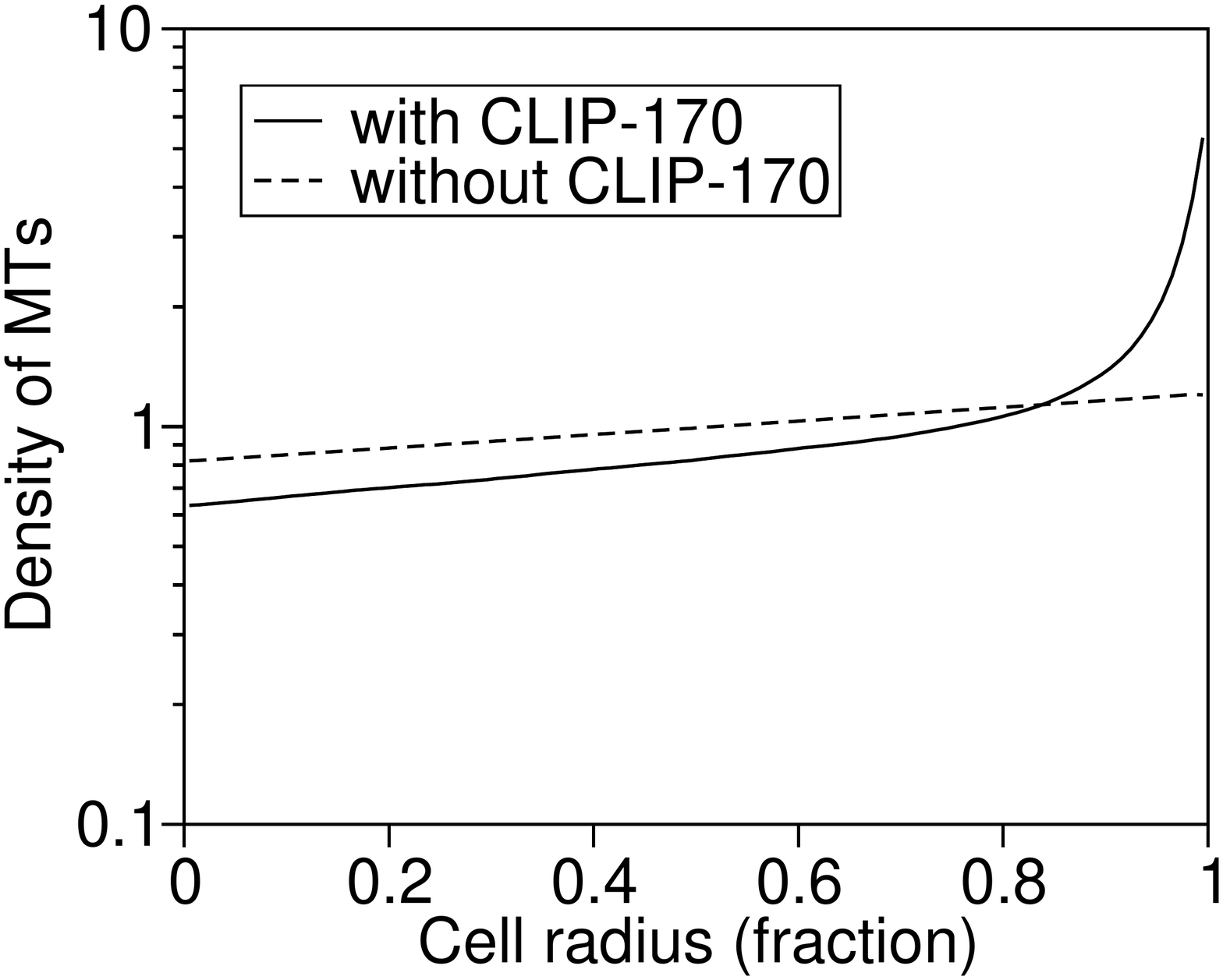}
  \end{center}
\textbf{Fig. S2. }Distribution of active plus ends along the cell radius with (solid line) and without (dashed line) CLIP-170. The data is the same that has been used to plot the histogram in Fig. 3. The plot is semi-logarithmic, thus indicating that the increase of MT plus ends towards the cell boundary is not part of an exponential distribution.
\end{figure}

\begin{figure}[h]
  \begin{center}
    \includegraphics[scale=0.25, clip]{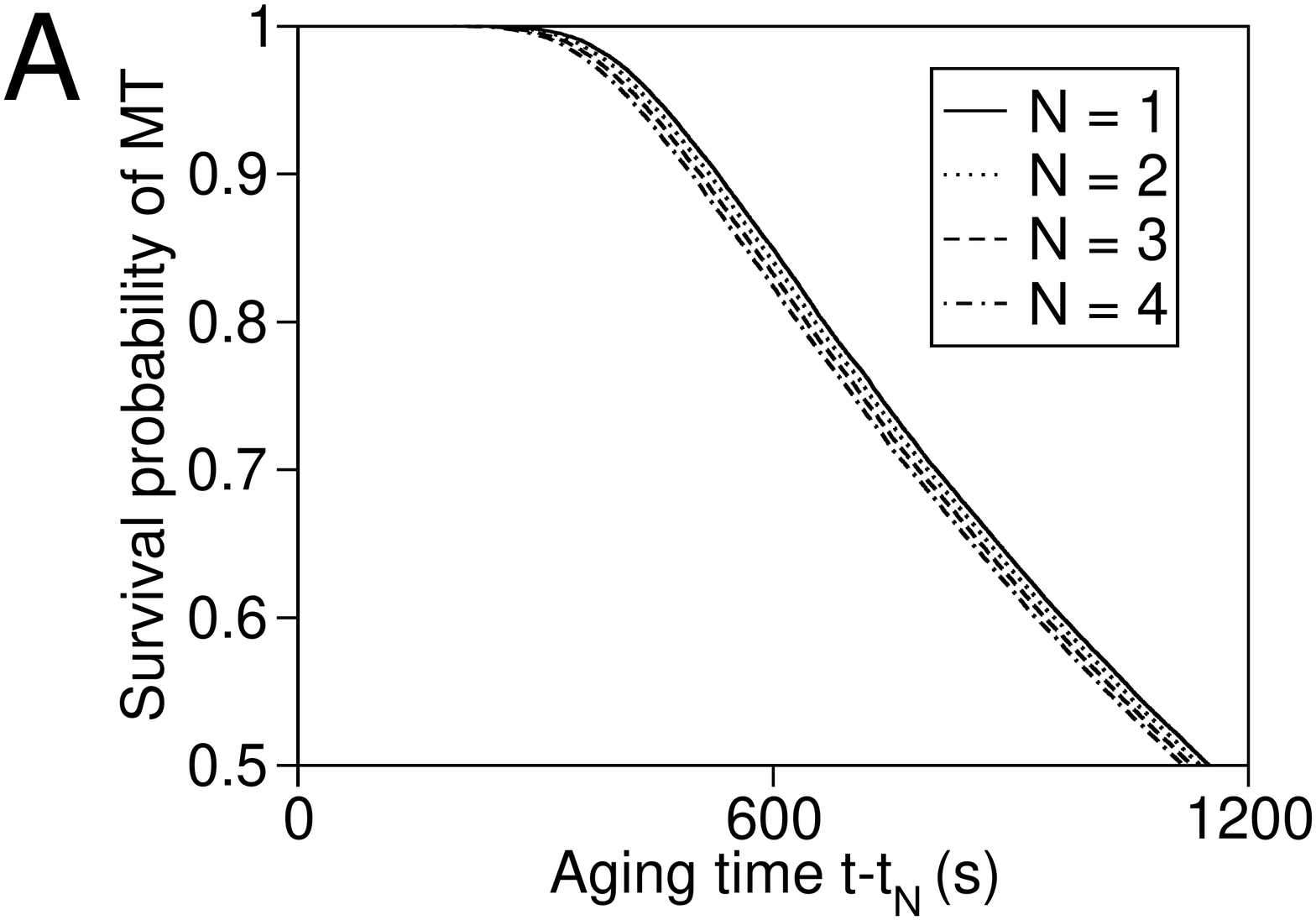}
    \includegraphics[scale=0.25, clip]{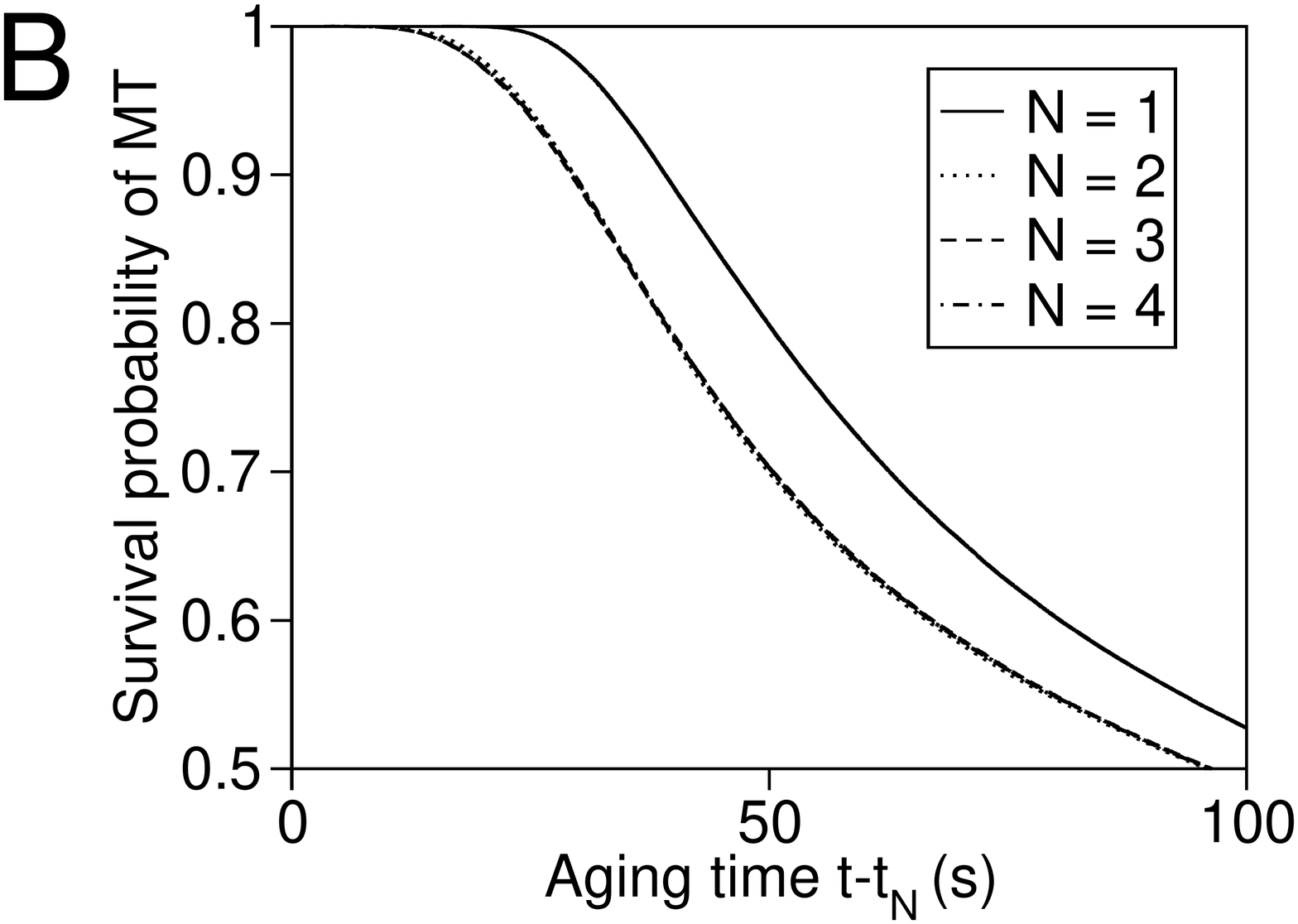}
    \includegraphics[scale=0.25, clip]{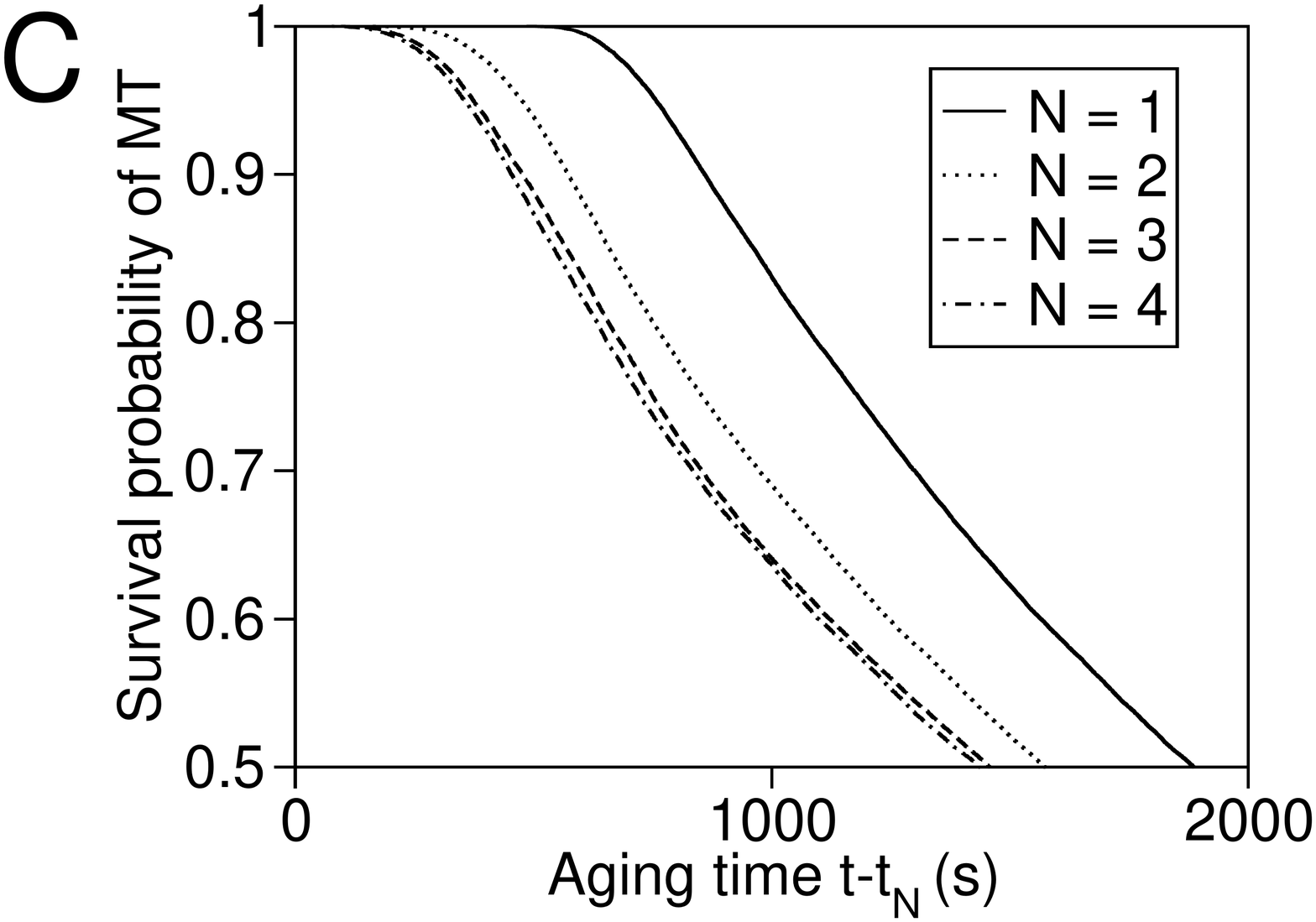}
    \includegraphics[scale=0.25, clip]{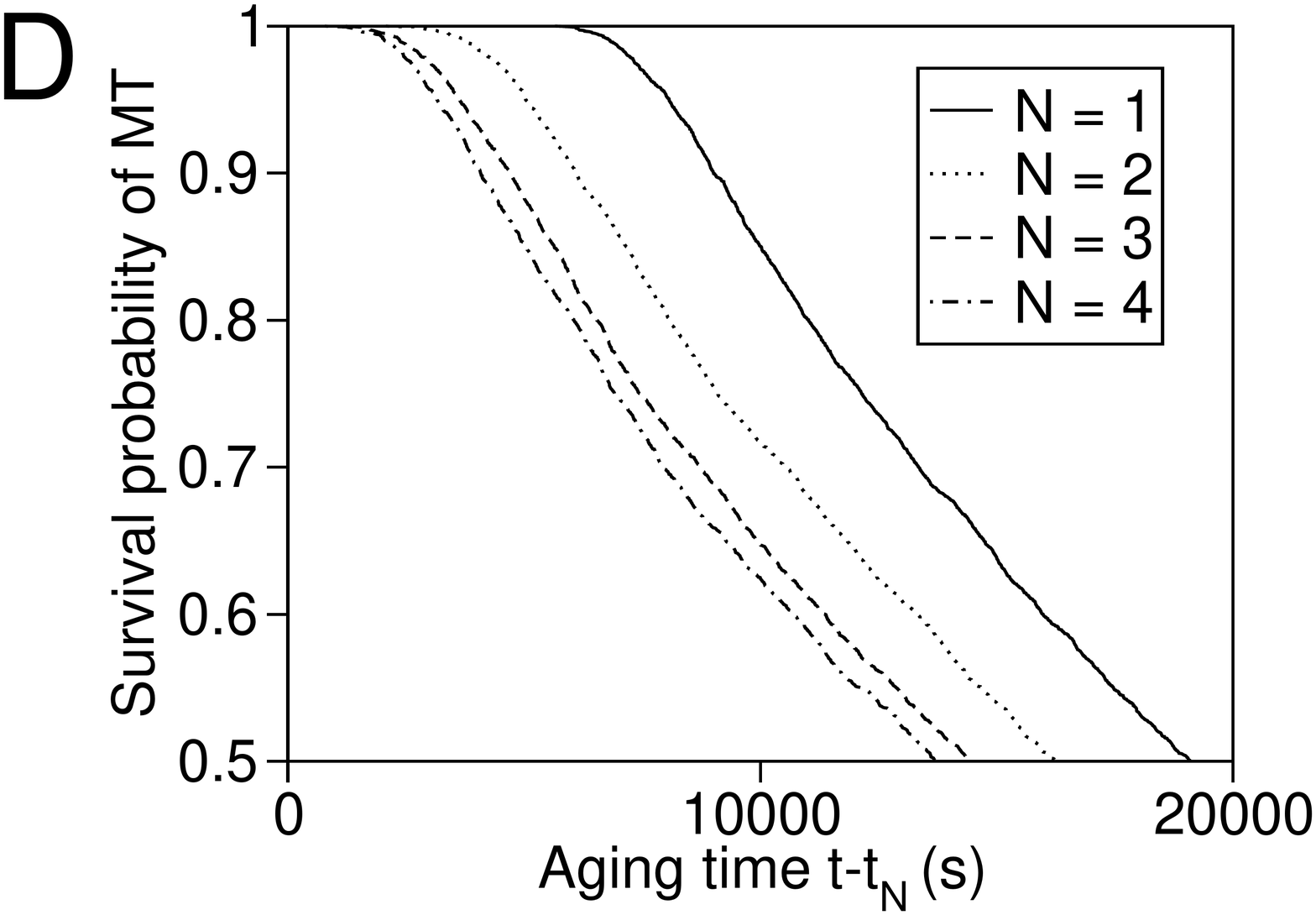}
    \includegraphics[scale=0.25, clip]{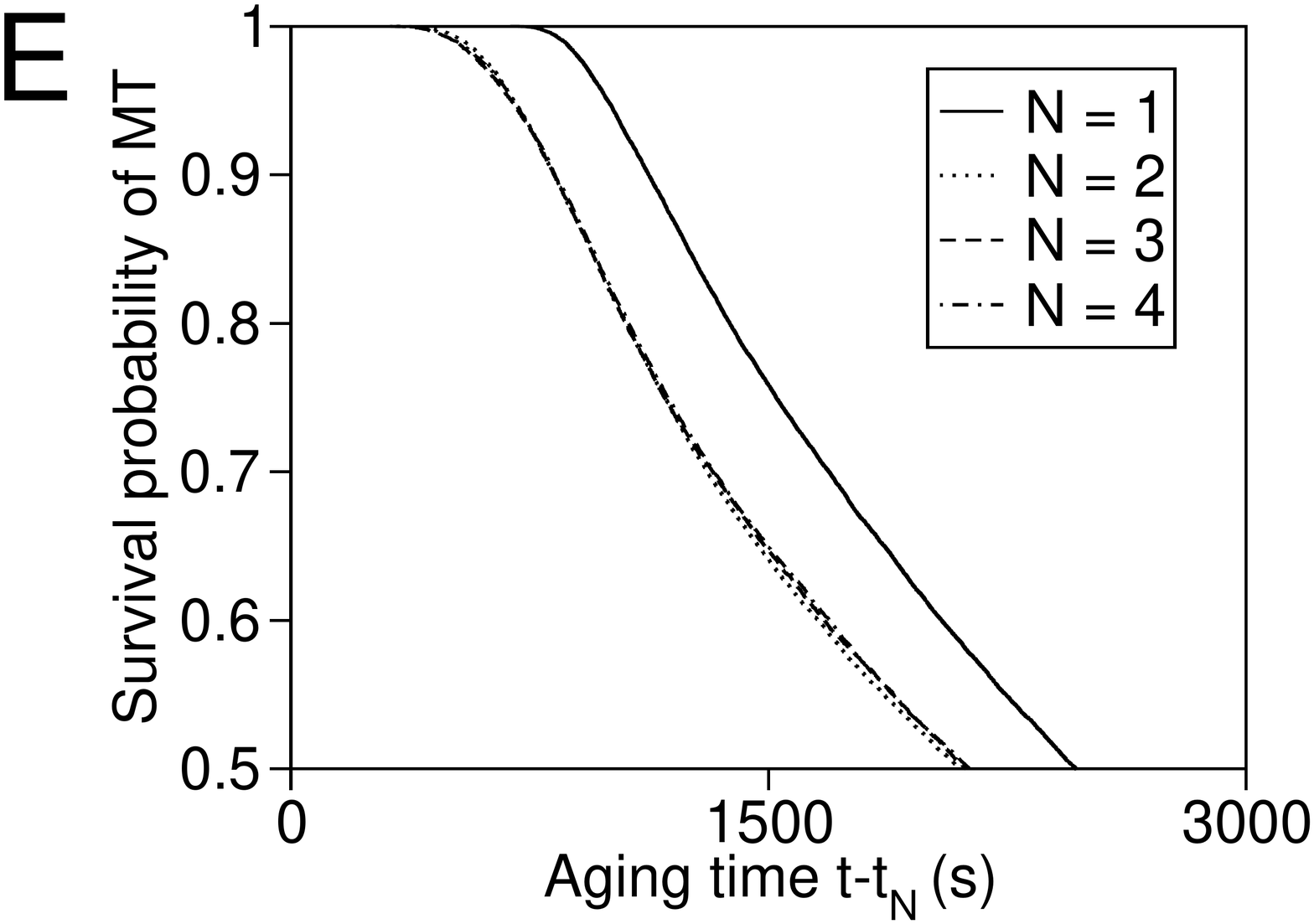}
    \includegraphics[scale=0.25, clip]{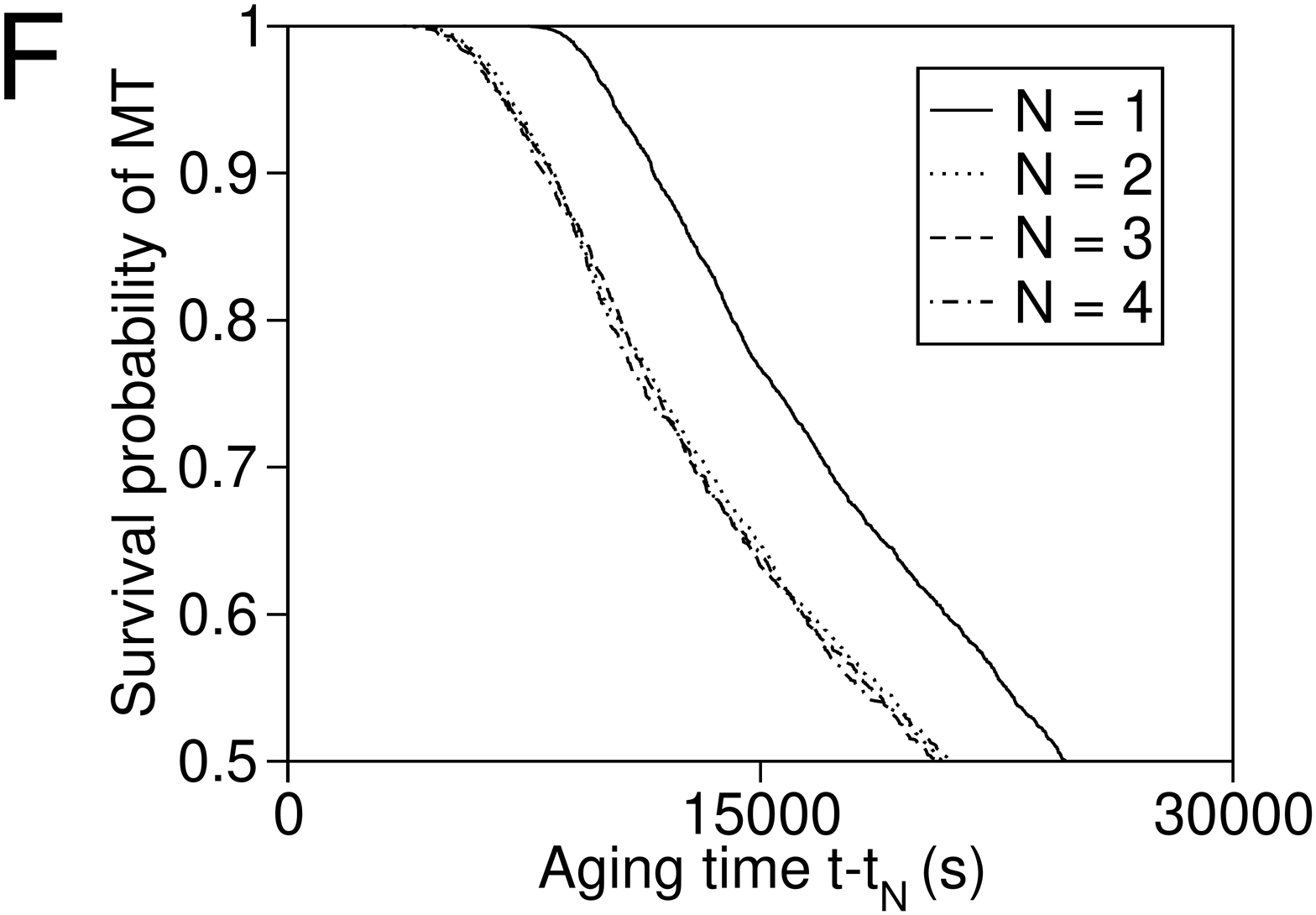}
  \end{center}
\textbf{Fig. S3. }Survival probability of MT after boundary contact with altered CLIP-170 dynamics. In order to better illustrate the effect of aging through dynamical stabilization, the same curves as in Fig. 4A are plotted for different parameters of the CLIP-170 dynamics: (A) $\tilde\nu_r=3.5~s^{-1}$, $\nu_d=10^{-2}~s^{-1}$; (B) $\tilde\nu_r=40~s^{-1}$, $\nu_d=1.5\cdot10^{-1}~s^{-1}$; (C) $\tilde\nu_r=40~s^{-1}$, $\nu_d=10^{-2}~s^{-1}$; (D) $\tilde\nu_r=40~s^{-1}$, $\nu_d=10^{-3}~s^{-1}$; (E) $\tilde\nu_r=400~s^{-1}$, $\nu_d=10^{-2}~s^{-1}$; (F) $\tilde\nu_r=400~s^{-1}$, $\nu_d=10^{-3}~s^{-1}$. Note the difference in time scales on the horizontal axis.
\end{figure}

\clearpage

\section*{Supplementary references}
\begin{tabbing}
S1. \quad\=W. Krauth, {\it Computations and Algorithms} (Oxford University Press, Oxford, 2006).\\
S2.\>Y. A. Komarova, A. S. Akhmanova, S. Kojimo, N. Galjart, G. G. Borisy, {\it J. Cell Biol.} {\bf 159}, \\
\>589 (2002).\\
S3.\>B. S. Govindan, J. W. B. Spillman, {\it Phys. Rev. E} {\bf 70}, 032901 (2004).\\
S4.\>Y. A. Komarova, I. A. Vorobjev, G. G. Borisy, {\it J. Cell Sci.} {\bf 115}, 3527 (2002).
\end{tabbing}

\end{document}